\documentclass[prl,
               preprint,
               superscriptaddress,
               onecolumn,
               showpacs,
               ]{revtex4-1}
               
\usepackage{graphicx}   
\usepackage{dcolumn}    
\usepackage{bm}         
\usepackage{amsmath}
\usepackage{amssymb}
\usepackage[usenames]{color}      

\newcommand{\Pra}{\mbox{\textit{Pr}}}  
\newcommand{\Ray}{\mbox{\textit{Ra}}}  
\newcommand{\Rey}{\mbox{\textit{Re}}}  
\newcommand{\Nus}{\mbox{\textit{Nu}}}  

\begin{document}

\title{Two-Scalar Turbulent Rayleigh-B\'{e}nard Convection: Numerical Simulations and Unifying Theory}

\author{Yantao Yang}
\affiliation{SKLTCS and Department of Mechanics and Engineering Science, College of Engineering, Peking University, Beijing 100871, China}
\affiliation{Physics of Fluids Group, Department of Science and Technology, Mesa+ Institute, Max Planck Center Twente for Complex Fluid Dynamics, and J. M. Burgers Center for Fluid Dynamics, University of Twente, 7500 AE Enschede, The Netherlands}
\author{Roberto Verzicco}
\affiliation{Physics of Fluids Group, Department of Science and Technology, Mesa+ Institute, Max Planck Center Twente for Complex Fluid Dynamics, and J. M. Burgers Center for Fluid Dynamics, University of Twente, 7500 AE Enschede, The Netherlands}
\affiliation{Dipartimento di Ingegneria Industriale, University of Rome ``Tor Vergata'', Via del Politecnico 1, Roma 00133, Italy}
\author{Detlef Lohse}
\affiliation{Physics of Fluids Group, Department of Science and Technology, Mesa+ Institute, Max Planck Center Twente for Complex Fluid Dynamics, and J. M. Burgers Center for Fluid Dynamics, University of Twente, 7500 AE Enschede, The Netherlands}

\date{\today}

\begin{abstract}
\noindent We conduct direct numerical simulations for turbulent Rayleigh-B\'{e}nard (RB) convection, driven simultaneously by two scalar components (say, temperature and salt concentration) with different molecular diffusivities, and measure the respective fluxes and the Reynolds number. To account for the results, we generalize the Grossmann-Lohse theory for traditional RB convections~(Grossmann and Lohse, {\it J. Fluid Mech.}, \textbf{407}, 27-56; {\it Phys. Rev. Lett.}, \textbf{86}, 3316-3319; Stevens {\it et al.}, {\it J. Fluid Mech.}, \textbf{730}, 295-308) to this two-scalar turbulent convection. Our numerical results suggest that the generalized theory can successfully predict the overall trends for the fluxes of two scalars and the Reynolds number. In fact, for most of the parameters explored here, the theory can even predict the absolute values of the fluxes and the Reynolds number with good accuracy. The current study extends the generality of the Grossmann-Lohse theory in the area of the buoyancy-driven convection flows.
\end{abstract}

\maketitle

\section{Introduction}

Rayleigh-B\'{e}nard (RB) convection serves as a commonly used system for studying natural convection which is ubiquitous in many nature and engineering environments. RB convection refers to a fluid layer which is heated from below and cooled from above, and is subject to an external gravitation field. Such systems have been extensively studied in recent years, e.g.~see the reviews of~\cite{Ahlers2009}, \cite{Lohse2010}, and \cite{Chilla2012}. One key question is how the scalar flux and flow velocity depend on the control parameters. The unifying theory for the flux and flow velocity~\citep[``GL theory''][]{GL2000,GL2001,GL2002,GL2004}, which are measured respectively by the Nusselt number $\Nus$ and the Reynolds number $\Rey$, has achieved great success for RB flows~\citep{Ahlers2009,GLrefit2013}, and now has predictive power for the absolute values of $\Nus$ and $\Rey$ for given control parameters, i.e. the Rayleigh number $\Ray$ and the Prandtl number $\Pra$.

However, in reality the convection flow can be much more complex than the idealized RB system, as in the case of external rotation~\citep[e.g.][]{King2009,Wei2015}, inhomogeneities of the top and bottom boundaries~\citep[e.g.][]{Wang2017,Bakhuis2018}, or wall roughness~\citep[e.g.][]{Shishkina2011,Salort2014,Xie2017,Zhu2017}. In this study we will investigate another type of complexity, i.e. RB convection driven by {\it two} different scalar components. Multiple-component convection is commonly encountered in nature. For instance, the density of seawater is mainly determined by temperature and salinity, and chemical reaction flows usually have more than one species. In the Ocean the vertical convection flow driven by both temperature and salinity gradients is usually referred to as double diffusive convection (DDC)~\citep{Turner1985,Radko2013}. Our previous study on DDC was confined in the so-called fingering regime, where the fluid layer experiences an unstable salinity gradient and stable temperature gradient~\citep{ddcjfm2015,ddcjfm2016}.

Here we will focus on the convection flow driven {\it simultaneously} by two scalar components with different molecular diffusivities. Our previous study showed that the original GL model can be used to describe the salinity transfer in fingering DDC flow~\citep{ddcjfm2015,ddcjfm2016}. The theory has also been applied to DDC in the diffusive regime, in which the fluid layer is subjected to an unstable temperature gradient and a stable salinity gradient~\citep{Hieronymus2016}. In this study the RB convection is driven by two scalar components which are both unstably stratified. Recall that the key idea of the GL theory is to divide both the momentum and thermal fields into their own boundary layer and bulk regions. Then scalings are developed for the two respective contributions to the respective dissipation rates, leading to $\Nus(\Ray,~\Pra)$ and $\Rey(\Ray,~\Pra)$. It is straightforward to generalize this type of argument to multiple scalar fields. We will validate this generalization of the GL theory by direct numerical simulations.

The paper is organized as follows. In section 2 we will provide the governing dynamical equations of the system and the details of our simulations. In section 3 we will present the generalization and application of the GL theory to the two-scalar RB convection. Finally section 4 concludes the paper. 

\section{Governing equations and numerical simulations}

The flow under consideration is incompressible and the density $\rho$ is determined by two scalar components, say temperature $\theta(\textbf{x},t)$ and concentration field $s(\textbf{x},t)$. The Oberbeck-Buossinesq approximation is adopted, i.e. the fluid density depends linearly on both scalars as $\rho(\theta, s)=\rho_0[1 - \beta_\theta \theta + \beta_s s]$. $\rho_0$ is a reference density, while $\theta$ and $s$ are the temperature and concentration relative to their respective reference values. $\beta_\zeta$ is the positive expansion coefficient respectively for temperature ($\zeta=\theta$) and concentration ($\zeta=s$). From now on the subscript $\zeta=\theta$ or $s$ stands for a quantity associated to scalar $\zeta$. The signs before the two terms indicate that the density is bigger for either lower temperature or higher concentration, which are the usually cases in practice.

The governing equations consist of the momentum equation and the advection-diffusion equations of two scalars, which read
\begin{subequations} \label{eq:drb}
\begin{eqnarray}
  && \partial_t u_i + u_j \partial_j u_i = 
       - \partial_i p + \nu \partial_j^2 u_i 
       + g \delta_{i3} (\beta_\theta \theta - \beta_s s), \label{eq:momem}  \\
  && \partial_t \theta + u_j \partial_j \theta = 
       \kappa_\theta \partial_j^2 \theta,   \label{eq:temper}  \\
  && \partial_t s + u_j \partial_j s =
       \kappa_s \partial_j^2 s. \label{eq:concen}
\end{eqnarray}
\end{subequations}
Here $u_i$ with $i=1,~2,~3$ are three velocity components, $p$ is the kinematic pressure, $\nu$ is the kinematic viscosity, $g$ is the gravitational acceleration, and $\kappa_\zeta$ is the molecular diffusivity, respectively. The dynamic system is further constrained by the continuity equation $\partial_i u_i=0$.

The fluid layer is between two parallel plates which are perpendicular to gravity and separated by a height $H$. At each plate both the temperature and concentration are kept constant, and the scalar difference between two plates is denoted by $\Delta$. The top plate has lower temperature and higher concentration, thus the flow is driven by both scalars. The flow has four control parameters, namely two Prandtl numbers and two Rayleigh numbers,
\begin{equation}
  \Pra_\zeta = \frac{\nu}{\kappa_\zeta}, \quad \quad 
  \Ray_\zeta = \frac{g \beta_\zeta \Delta_\zeta H^3}{\kappa_\zeta \nu}.
\end{equation}
Another useful parameter, which is borrowed from the DDC community, is the density ratio
\begin{equation}
  \Lambda = \frac{\beta_\theta \Delta_\theta}{\beta_s \Delta_s} = \frac{\Pra_s \Ray_\theta}{\Pra_\theta \Ray_s}.
\end{equation}
$\Lambda$ measures the relative strength of the buoyancy force induced by the temperature difference to that induced by the concentration difference. $\Lambda<1$ indicates that the buoyancy force of the concentration difference is stronger than that of the temperature field, which we refer to as the ``concentration-dominant'' (CD) regime. Accordingly, $\Lambda>1$ is referred to as the ``temperature-dominant'' (TD) regime. Three key responses of the system are the scalar fluxes and the flow velocity, which are measured by the two Nusselt numbers and the Reynolds number
\begin{equation}\label{eq:Nuss}
  \Nus_s = \frac{\langle u_3 s \rangle - \kappa_s \partial_3\langle s \rangle}{\kappa_s \Delta_s H^{-1}},
 \quad\quad
  \Nus_\theta = \frac{\langle u_3 \theta \rangle - \kappa_\theta \partial_3\langle \theta \rangle}{\kappa_\theta \Delta_\theta H^{-1}},
 \quad\quad 
  \Rey = \frac{u_{rms} H}{\nu}.
\end{equation}
Here $\langle\cdot\rangle$ represents the spatial and time average. $u_{rms}$ is the root-mean-square value of the velocity magnitude. 

The governing equation (\ref{eq:drb}) is numerically solved by using a highly efficient code developed in our group~\citep{multigrid2015}, which has been used intensively in our previous DDC studies~\citep{ddcjfm2015,ddcjfm2016}. The code employs a multiple-grid method, which solves the momentum and fast-diffusing scalar on a base mesh, and the slow diffusing scalar on a refined mesh, respectively. The resolution is chosen to meet the criteria proposed in~\cite{Stevens2010}. The flow quantities are non-dimensionalized by the height $H$, the free-fall velocity defined by concentration difference $U_c=\sqrt{g\beta_s\Delta_sH}$, and the scalar differences $\Delta_\zeta$, respectively. At two plates no-slip boundary conditions are applied and both scalars are kept constant. In the two horizontal directions the periodic boundary conditions are employed. We fix $\Pra_\theta=1$ and change $\Ray_\theta$, $\Ray_s$, and $\Pra_s$. In this work we always set $\Pra_s > \Pra_\theta$, which implies that concentration diffuses slower than temperature. The simulated cases are summarized in table~\ref{tab:nume}. All cases are sorted into four groups, as shown in table~\ref{tab:nume}. Within each group we only vary one control parameter and fix all others constant.
\begin{table}
\begin{center}
\setlength{\tabcolsep}{8pt}
\begin{tabular}{cccccccccc}
  $\Pra_s$ & $\Ray_\theta$ & $\Ray_s$ & $\Lambda$ & $\Gamma$ & $n_x$ ($m_x$) & $n_z$ ($m_z$) & $\Nus_\theta$ & $\Nus_s$ & $\Rey$ \\[0.2cm]
  10.0   & $10^6$  & $10^5$  & 100.0  & 4.0 & 240(3) & 128(2) & 18.481  & 8.2880  & 221.24  \\
  10.0   & $10^6$  & $10^6$  & 10.0   & 4.0 & 240(3) & 128(2) & 18.737  & 8.3676  & 223.03  \\
  10.0   & $10^6$  & $10^7$  & 1.0    & 4.0 & 240(4) & 192(2) & 20.809  & 8.9735  & 238.63  \\
  10.0   & $10^6$  & $10^8$  & 0.1    & 4.0 & 288(4) & 192(2) & 31.683  & 12.456  & 395.77  \\
  10.0   & $10^6$  & $10^9$  & 0.01   & 4.0 & 512(4) & 384(2) & 60.811  & 22.773  & 1083.3  \\[0.2cm]
    
  10.0   & $10^7$  & $10^5$  & 1000.0 & 4.0 & 384(4) & 192(2) & 35.609  & 15.769  & 681.74  \\
  10.0   & $10^7$  & $10^6$  & 100.0  & 4.0 & 384(4) & 192(2) & 35.723  & 15.841  & 680.93  \\
  10.0   & $10^7$  & $10^7$  & 10.0   & 4.0 & 384(4) & 192(2) & 36.131  & 15.946  & 687.62  \\
  10.0   & $10^7$  & $10^8$  & 1.0    & 4.0 & 384(4) & 192(2) & 40.168  & 17.152  & 744.13  \\
  10.0   & $10^7$  & $10^9$  & 0.1    & 2.0 & 288(4) & 288(2) & 63.993  & 24.929  & 1208.2  \\[0.2cm]
  
  10.0   & $10^4$  & $10^7$  & 0.01   & 4.0 & 256(2) & 192(1) & 16.289  & 6.0401  & 99.597  \\
  10.0   & $10^5$  & $10^7$  & 0.1    & 4.0 & 256(2) & 192(1) & 16.714  & 6.5328  & 119.79  \\
  10.0   & $10^6$  & $10^7$  & 1.0    & 4.0 & 240(4) & 192(2) & 20.809  & 8.9735  & 238.63  \\
  10.0   & $10^7$  & $10^7$  & 10.0   & 4.0 & 384(4) & 192(2) & 36.131  & 15.946  & 687.62  \\
  10.0   & $10^8$  & $10^7$  & 100.0  & 2.0 & 384(4) & 384(2) & 72.405  & 31.514  & 1937.7  \\[0.2cm]
  
  1.0    & $10^7$  & $10^7$  & 1.0    & 4.0 & 384(1) & 192(1) & 19.341  & 19.341  & 949.51  \\
  2.0    & $10^7$  & $10^7$  & 2.0    & 4.0 & 384(2) & 192(1) & 22.759  & 17.340  & 779.04  \\
  5.0    & $10^7$  & $10^7$  & 5.0    & 4.0 & 384(3) & 192(2) & 29.362  & 16.229  & 705.87  \\
  10.0   & $10^7$  & $10^7$  & 10.0   & 4.0 & 384(4) & 192(2) & 36.131  & 15.946  & 687.62  \\
  30.0   & $10^7$  & $10^7$  & 30.0   & 2.0 & 384(6) & 385(2) & 52.525  & 16.503  & 692.06  \\
\end{tabular}
\end{center}
\caption{Summary of the control parameters, numerical details, and the global responses. For all cases $\Pra_\theta=1$. Columns from left to right: Prandtl number of concentration field, Rayleigh numbers of temperature and concentration, density ratio, aspect ratio of domain (horizontal length over height), horizontal resolution of base mesh (refinement factor of refined mesh), vertical resolution of base mesh (refinement factor of refined mesh), two Nusselt numbers, and Reynolds number. Note that some cases appear repeatedly for the completeness of each group.}
\label{tab:nume}
\end{table}%

\begin{figure}
\centering
\includegraphics[width=\textwidth]{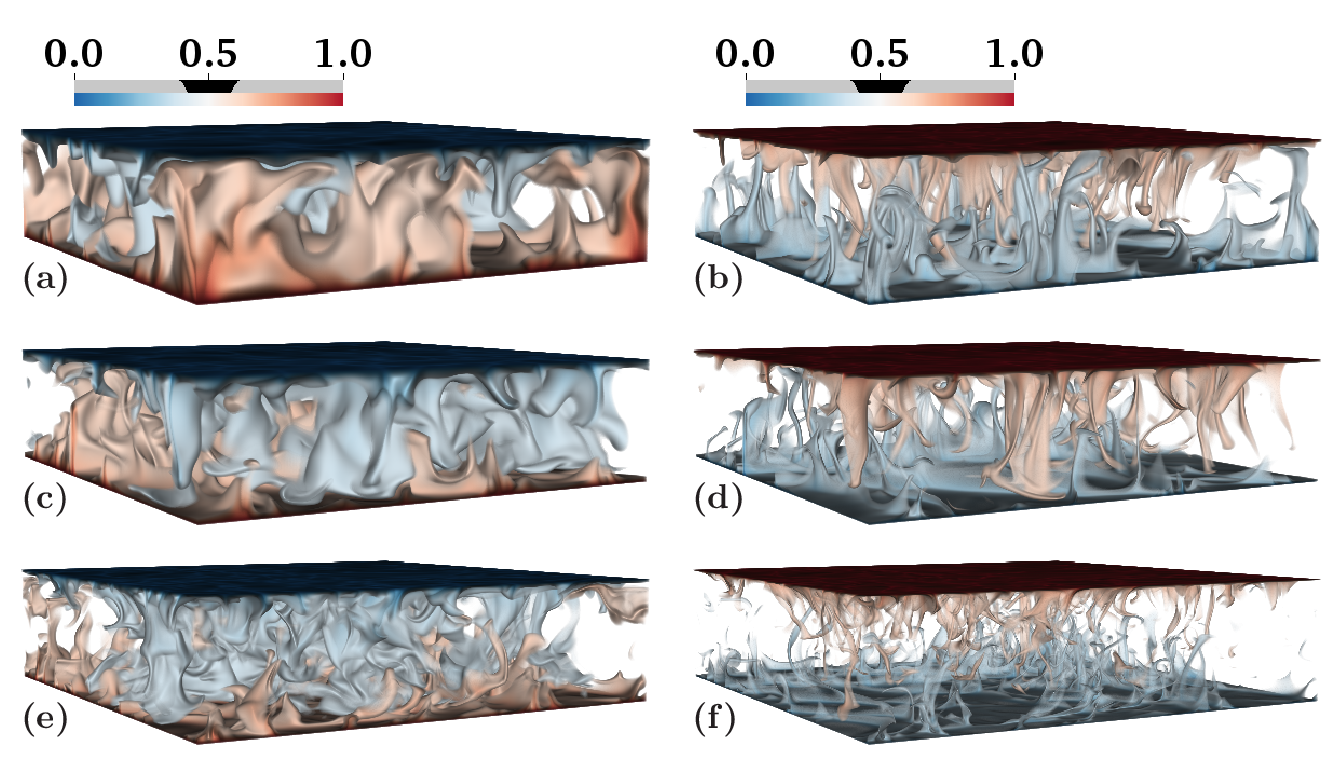}%
\caption{Three-dimensional volume rendering of the temperature (left column) and the concentration (right column) for three different runs with $\Pra_s=10$ and $\Ray_s=10^7$. From top row to bottom $\Ray_\theta=10^5$, $10^6$, and $10^7$, or equivalently $\Lambda=0.1$, $1$, and $10$, respectively.}
\label{fig:field}
\end{figure}

In figure~\ref{fig:field} we show the scalar fields of three different runs with fixed $\Pra_s=10$ and $\Ray_s=10^7$ but different $\Ray_\theta=10^5$ (a, b), $10^6$ (c, d), and $10^7$ (e, f), or equivalently $\Lambda=0.1$, $1$, and $10$, respectively. For the case shown in figures~\ref{fig:field}(a,~b) with $\Lambda=0.1$ the buoyancy force is dominated by the concentration difference. Since for this case the molecular diffusivity of temperature is ten times faster than that of concentration, the typical size of the temperature plumes is much larger than the concentration ones due to the fast horizontal diffusion. On the contrary, for $\Lambda=10$ as shown in figures~\ref{fig:field}(e, f) the temperature component contributes the most part of the buoyancy force, and the temperature plumes are very active. Now the concentration plumes become thin filaments embedded within temperature plumes. When $\Lambda=1$ (figures~\ref{fig:field}c, d) the two scalar components contribute equally to the total buoyancy force, and the size of the scalar plumes is in between of the other two cases. 

\begin{figure}
\centering
\includegraphics[width=\textwidth]{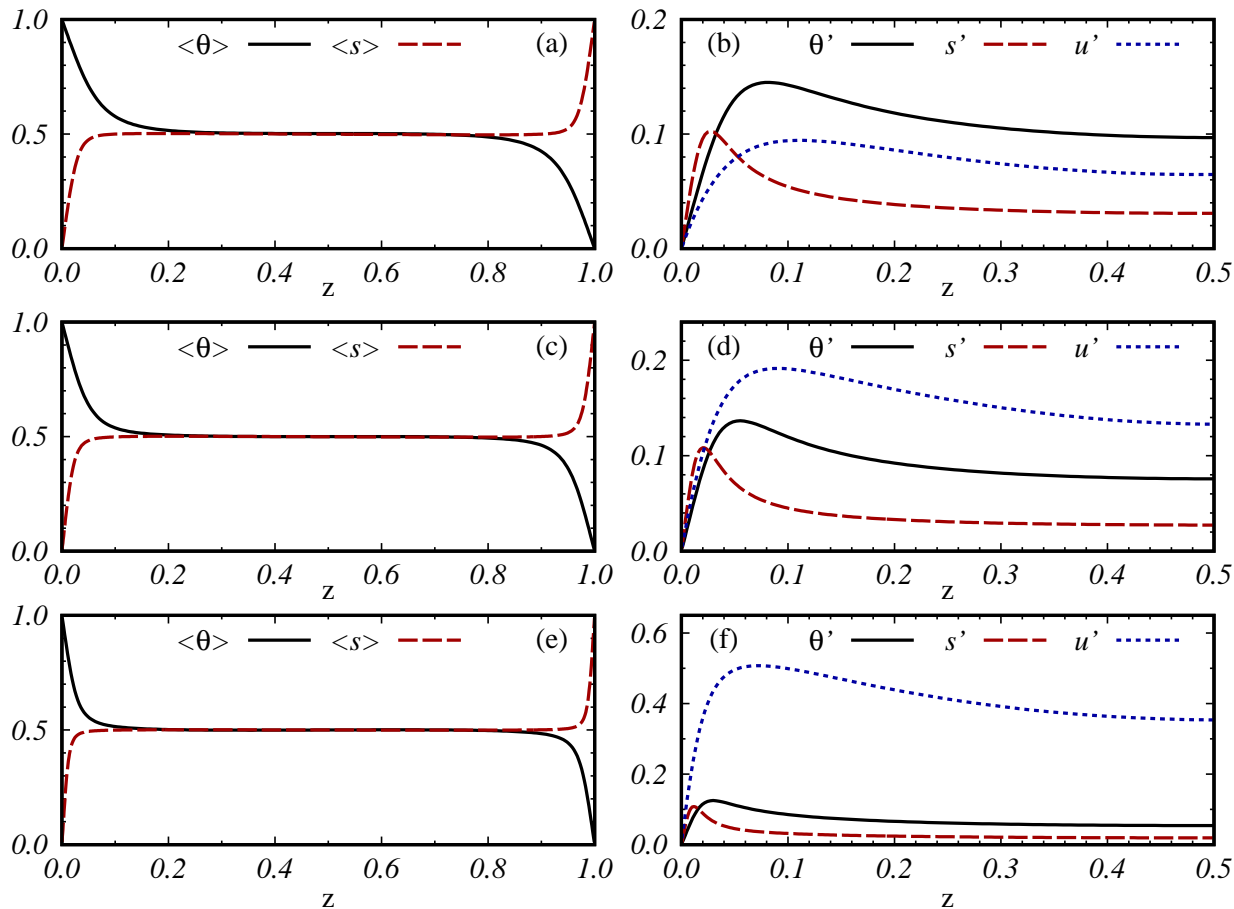}%
\caption{Mean profiles of two scalars (left column) and the root-mean-square of scalars and one horizontal velocity coponent (right column) for the three cases shown in figure~\ref{fig:field} (from top to bottom). For clarity of the near wall region and due to the symmetry about $z=0.5$, in the right column we only show the lower half of the domain.}
\label{fig:profs}
\end{figure}

Figure~\ref{fig:profs} displays the profiles of the flow quantities for the three cases shown in figure~\ref{fig:field}. The mean profiles of the scalars suggest that both scalar fields have two thin boundary layer regions with high gradient adjacent to the plates, and in between a well mixed bulk region with nearly constant mean values (see figures~\ref{fig:profs}a,~\ref{fig:profs}c,~\ref{fig:profs}e). In figures~\ref{fig:profs}b,~\ref{fig:profs}d, and~\ref{fig:profs}f we plot the root-mean-square (rms) profiles of the fluctuations of scalars and one horizontal velocity component near the bottom plate. As in the RB flow, the peak location in rms profile can be regarded as the height of the boundary layers. Note that for the three cases shown here, $\Ray_s$ is fixed at $10^7$, and $\Ray_\theta$ increases from $10^5$ to $10^7$. As $\Ray_\theta$ becomes larger, the total driving buoyancy force increases. The velocity fluctuation becomes stronger, indicating more intense turbulence. Accordingly, the peak location of velocity rms moves closer to the plate. Meanwhile, the boundary layer thickness of both scalars decreases and the concentration boundary layer is always nested inside the temperature boundary layer due to its smaller diffusivity than that of temperature.

Interestingly, the momentum boundary layer has larger thickness than both scalar fields, even for the case with $\Ray_\theta=10^5$ shown in figure~\ref{fig:profs}b. Note that for this case $\Lambda=0.1$, meaning that the buoyancy force of the temperature difference is much smaller than that of the concentration difference. But the momentum boundary layer thickness is still set by the temperature field. This is consistent with the flow structures shown in figure~\ref{fig:field}, since temperature diffuses faster than concentration and its boundary layer extends beyond that of concentration field. Thermal plumes grow from a location higher than concentration plumes do and therefore the temperature field sets the momentum boundary layer height even for very small $\Lambda$.

\section{Generalized Grossmann-Lohse theory}

The GL theory was originally developed for RB flow~\citep{GL2000, GL2001, GL2002, GL2004} to provide a unifying theory for $\Nus(\Ray,~\Pra)$ and $\Rey(\Ray,~\Pra)$, and successfully accounts for most of the existing experimental and numerical results~\citep{Ahlers2009, GLrefit2013}. Our previous DDC studies~\citep{ddcjfm2015,ddcjfm2016} showed that the original theory, without any modification of the coefficients, can also be used to describe the concentration flux and flow velocity scalings of the fingering DDC flow, i.e. flow driven by a concentration difference and stablized by a temperature difference. In this section we will briefly discuss the theory and the formulations, and then apply it to the current two-scalar convection flow.

The theory is built upon the exact relations between the dissipation rates and the global fluxes, which for the present problem read
\begin{subequations}\label{eq:dsp}
\begin{eqnarray}
   \epsilon_\theta &\equiv& \left\langle \kappa_\theta [\partial_i \theta]^2 \right\rangle_V  
                  = \kappa_\theta\, (\Delta_\theta)^2\, H^{-2}\, \Nus_\theta,   \label{eq:disst}  \\
   \epsilon_s &\equiv& \left\langle \kappa_s [\partial_i s]^2 \right\rangle_V 
                  = \kappa_s\, (\Delta_s)^2\, H^{-2}\, \Nus_s,  \label{eq:disss} \\ 
   \epsilon_u &\equiv& \left\langle \nu[\partial_i u_j]^2 \right\rangle_V 
                  = \nu^3 H^{-4}\, \left[ \Ray_\theta\, \Pra^{-2}_\theta (\Nus_\theta - 1) 
                        + \Ray_s\, \Pra^{-2}_s (\Nus_s - 1) \right]. \label{eq:disse}
\end{eqnarray}
\end{subequations}
The flow domain is divided into the boundary-layer and bulk regions. For each region different scalings are derived for the individual dissipation rates. Such division is still applicable to the current flow, as shown in figure~\ref{fig:field}. Following the same arguments as in the original theory, one finally obtains the GL theory for the turbulent two-scalar convective flow, namely
\begin{subequations}\label{eq:gl}
\begin{eqnarray}
 && (\Nus_s - 1) \Ray_s \Pra_s^{-2} + (\Nus_\theta - 1) \Ray_\theta \Pra_\theta^{-2}
           = c_1 \frac{\Rey^2}{g\left(\sqrt{\Rey_c/\Rey}\right)} + c_2 \Rey^3, \label{eq:gl-a} \\[0.2cm]
 && \Nus_\theta  - 1 = c_{3}  \Rey^{1/2}\,\Pra_\theta ^{1/2} 
           \left\{ f\left[ \frac{2a \Nus_\theta  }{\sqrt{\Rey_c}}
           g\left( \sqrt{\frac{\Rey_c}{\Rey}} \right) \right] \right\}^{1/2} \nonumber\\[0.2cm]
 && \hspace{4cm}    + c_{4} \Rey\, \Pra_\theta\, f\left[ \frac{2a \Nus_\theta }{\sqrt{\Rey_c}}
                               g\left( \sqrt{\frac{\Rey_c}{\Rey}} \right) \right], \label{eq:gl-b} \\ 
 && \Nus_s - 1 = c_{3} \Rey^{1/2}\,\Pra_s^{1/2} \left\{ f\left[ \frac{2a\Nus_s}{\sqrt{\Rey_c}}
               g\left( \sqrt{\frac{\Rey_c}{\Rey}} \right) \right] \right\}^{1/2} \nonumber\\[0.2cm]
 && \hspace{4cm}    + c_{4} \Rey\,\Pra_s\, f\left[ \frac{2a\Nus_s}{\sqrt{\Rey_c}}
                               g\left( \sqrt{\frac{\Rey_c}{\Rey}} \right) \right]. \label{eq:gl-c} 
\end{eqnarray}
\end{subequations}
Note that as explained in~\cite{ddcjfm2015}, when either of the two scalar differences decreases to zero, the flow will reduce to the traditional RB flow with one scalar and the theory should degenerate to the original form. Thus there are only the five original coefficients $c_i$ with $i=1,2,3,4$ and $a$ in~(\ref{eq:gl-b}) and (\ref{eq:gl-c}), and not two extra ones for the second scalar field, as one may naively assume. The $c_i$ with $i=1,2,3,4$ and $a$ can directly be taken from~\cite{GLrefit2013}, i.e.
\begin{equation}\label{eq:coefs}
  c_1=8.05, \quad c_2=1.38, \quad c_3=0.487, \quad c_4=0.0252, \quad a=0.922.
\end{equation}

\begin{figure}
\begin{center}
   \includegraphics[width=\textwidth]{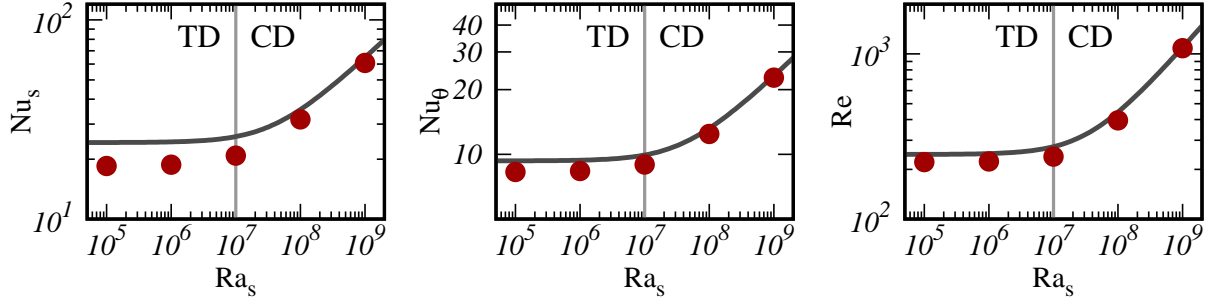}%
   \\
   (a) $\Pra_s=10$ and $\Ray_\theta=10^6$ \\[0.1cm]
   \includegraphics[width=\textwidth]{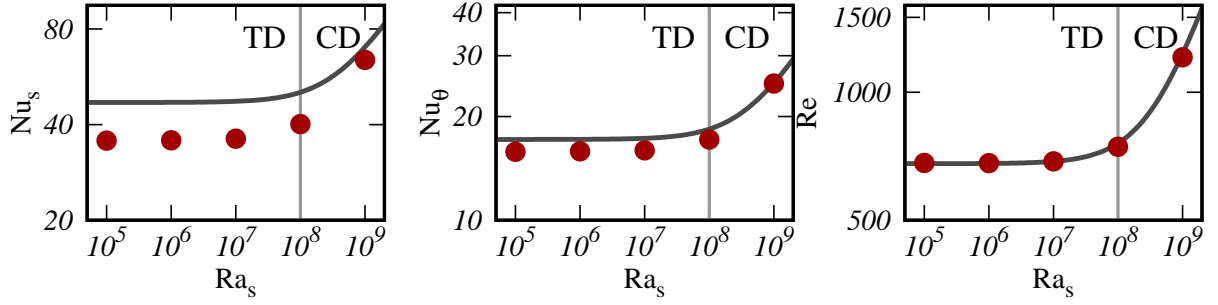}%
   \\
   (b) $\Pra_s=10$ and $\Ray_\theta=10^7$ \\[0.1cm]
   \includegraphics[width=\textwidth]{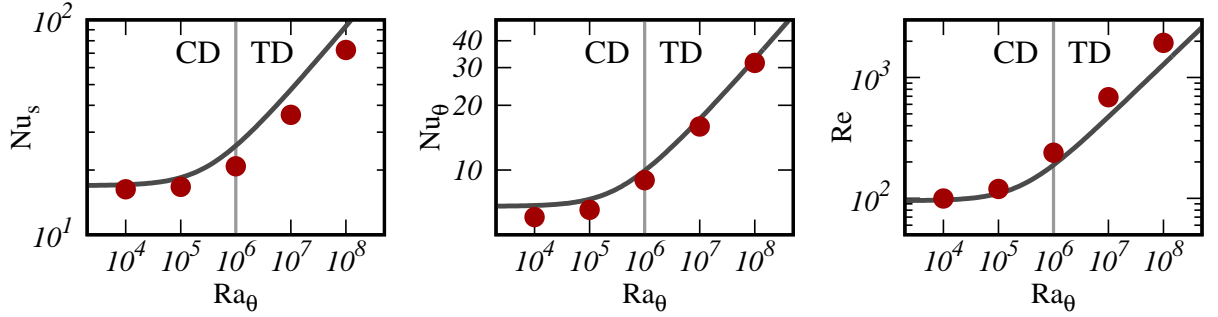}%
   \\
   (c) $\Pra_s=10$ and $\Ray_s=10^7$ \\[0.1cm]
   \includegraphics[width=\textwidth]{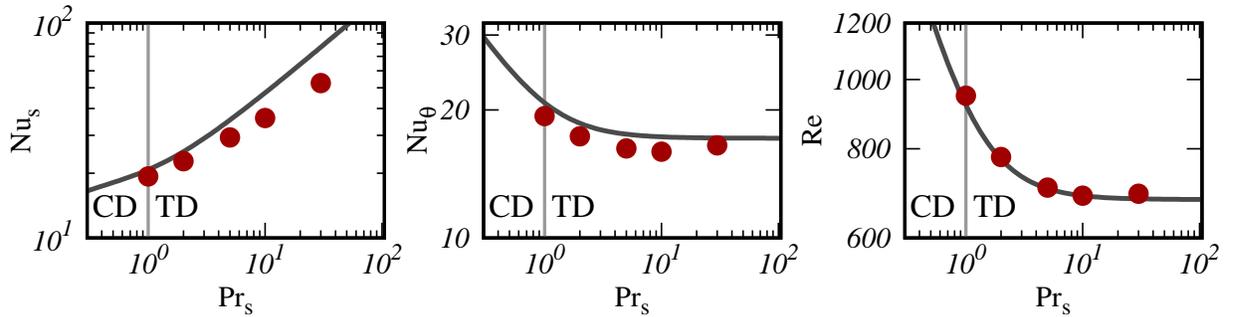}%
   \\
   (d) $\Ray_\theta=10^7$ and $\Ray_s=10^7$\\[0.1cm]
\caption{Comparison of the GL theory (lines) to the numerical results (symbols). Each subfigure shows a group of cases listed in table~\ref{tab:nume}. The vertical gray line in every plot indicates the location $\Lambda=1$, which separates the CD and TD regimes.}
\label{fig:glfit}
\end{center}
\end{figure}

We now compare the theory to the current numerical results. To correctly predict the Reynolds number a transformation coefficient $\alpha$ needs to be determined and the procedure is described in~\cite{GLrefit2013}. Here we fix $\alpha=1.453$ by using the Reynolds number of the case with $\Pra_s=10$, $\Ray_\theta=\Ray_s=10^7$. Figure~\ref{fig:glfit} shows both the theoretical predictions and the numerical measurements for the four groups of the cases listed in table~\ref{tab:nume}. The overall trends of all three global responses, namely $\Nus_\theta$, $\Nus_s$, and $\Rey$, are very well captured by the theory. Moreover, the theory can even predict the absolute values for most of the cases. 

However, some discrepancy is observed for certain parameters. Among the three global responses, the theoretical prediction for the concentration flux $\Nus_s$ exhibits the biggest deviation from the numerical results, especially in the deep TD regime (with large $\Lambda$). In this regime the flow is mainly driven by the temperature difference, such as the case shown in figures~\ref{fig:field}e and \ref{fig:field}f. All the concentration plumes are very thin and stay in the core regions of the temperature plumes, therefore the buoyancy anomaly associated to concentration and its effects on the momentum field are confined inside the temperature plumes due to the slow sideward diffusion. The morphology and dynamics of such thin concentration plumes are very different from the traditional RB plumes, thus the original scaling arguments of the GL theory become less accurate. 

For the opposite situation in the deep CD regime (with small $\Lambda$), although the buoyancy force is mainly generated by the concentration component, the temperature anomaly diffuses faster in the lateral direction and is not confined inside the concentration plumes. The temperature anomaly can directly interact with the momentum field and thus more similar to the situation in the RB flows. Therefore the GL theory performs better in the CD regime than in the TD regime, as shown in figures~\ref{fig:glfit}a-c. 

\section{Conclusions and discussions}

In summary, we conducted the direction numerical simulations of the RB convection driven by two scalar components which have different molecular diffusivities. The flow morphology changes for different ratios of the buoyancy forces associated to the two scalar differences. We have generalized the GL theory for the RB convection to the current problem. The results show that the theory captures the overall trends of the dependences of scalar fluxes and flow velocity on the control parameters. For most of the cases the theory predicts the absolute values with good accuracy. This comparison demonstrates the applicability of the GL theory to multiple component convection flows. The accuracy of the theory decreases when the fast-diffusing component dominates the buoyancy force, i.e. in the TD regime. We argue that in this regime, the structures of the slow-diffusing component is very different from those in the traditional RB flows, and the argument in the original GL theory becomes less accurate.

A more refined generalization of the GL theory should assume that the coefficients are not constant but some functions of the density ratio $\Lambda$, and they recover the original GL values when the system degenerates to a single scalar RB flows. The theory also deserves to be tested over a larger parameter space, such as for the scalars with Prandtl number smaller than one, which is of relevance to some astrophysical flows.

\section*{Acknowledgements}

This study is supported by the Dutch Foundation for Fundamental Research on Matter (FOM), and by the Netherlands Center for Multiscale Catalytic Energy Conversion (MCEC), an NWO Gravitation programme funded by the Ministry of Education, Culture and Science of the government of the Netherlands. The simulations were conducted on the Cartesius supercomputer at the Dutch national e-infrastructure of SURFsara, and the Marconi supercomputer based in CINECA, Italy through the PRACE project NO.~2016143351. 

\bibliographystyle{jfm}

\begin{thebibliography}{23}
\expandafter\ifx\csname natexlab\endcsname\relax\def\natexlab#1{#1}\fi
\def\au#1{#1} \def\ed#1{#1} \def\yr#1{#1}\def\at#1{#1}\def\jt#1{\textit{#1}}
  \def\bt#1{#1}\def\bvol#1{\textbf{#1}} \def\vol#1{#1} \def\pg#1{#1}
  \def\publ#1{#1}\def\arxiv#1{#1}\def\org#1{#1}\def\st#1{\textit{#1}}

\bibitem[Ahlers {\em et~al.\/}(2009)Ahlers, Grossmann \& Lohse]{Ahlers2009}
{\sc \au{Ahlers, G.}, \au{Grossmann, S.} \& \au{Lohse, D.}} \yr{2009}  \at{Heat
  transfer and large scale dynamics in turbulent {Rayleigh-B\'enard}
  convection}.  \jt{Rev. Mod. Phys.}  \bvol{81},  \pg{503--537}.

\bibitem[Bakhuis {\em et~al.\/}(2018)Bakhuis, Ostilla-M\'{o}nico, van~der Poel,
  Verzicco \& Lohse]{Bakhuis2018}
{\sc \au{Bakhuis, D.}, \au{Ostilla-M\'{o}nico, R.}, \au{van~der Poel, E.P.},
  \au{Verzicco, R.} \& \au{Lohse, D.}} \yr{2018}  \at{Mixed insulating and
  conducting thermal boundary conditions in {Rayleigh-B\'{e}nard} convection}.
  \jt{J. Fluid Mech.}  \bvol{835},  \pg{491--511}.

\bibitem[Chill\`{a} \& Schumacher(2012)]{Chilla2012}
{\sc \au{Chill\`{a}, F.} \& \au{Schumacher, J.}} \yr{2012}  \at{New
  perspectives in turbulent {Rayleigh?B\'{e}nard} convection.}  \jt{Eur. Phys.
  J. E}  \bvol{35}~(7),  \pg{58}.

\bibitem[Grossmann \& Lohse(2000)]{GL2000}
{\sc \au{Grossmann, S.} \& \au{Lohse, D.}} \yr{2000}  \at{Scaling in thermal
  convection: a unifying theory}.  \jt{J. Fluid Mech.}  \bvol{407},
  \pg{27--56}.

\bibitem[Grossmann \& Lohse(2001)]{GL2001}
{\sc \au{Grossmann, S.} \& \au{Lohse, D.}} \yr{2001}  \at{Thermal convection
  for large {P}randtl numbers}.  \jt{Phys. Rev. Lett.}  \bvol{86}~(15),
  \pg{3316--3319}.

\bibitem[Grossmann \& Lohse(2002)]{GL2002}
{\sc \au{Grossmann, S.} \& \au{Lohse, D.}} \yr{2002}  \at{{Prandtl and
  Rayleigh} number dependence of the {Reynolds} number in turbulent thermal
  convection}.  \jt{Phys. Rev. E}  \bvol{66}~(1),  \pg{016305}.

\bibitem[Grossmann \& Lohse(2004)]{GL2004}
{\sc \au{Grossmann, S.} \& \au{Lohse, D.}} \yr{2004}  \at{Fluctuations in
  turbulent {Rayleigh-B\'{e}nard} convection: {T}he role of plumes}.  \jt{Phys.
  Fluids}  \bvol{16}~(12),  \pg{4462--4472}.

\bibitem[Hieronymus \& Carpenter(2016)]{Hieronymus2016}
{\sc \au{Hieronymus, M.} \& \au{Carpenter, J.R.}} \yr{2016}  \at{Energy and
  variance budgets of a diffusive staircase with implications for heat flux
  scaling}.  \jt{J. Phys. Oceanogr.}  \bvol{46},  \pg{2553--2569}.

\bibitem[King {\em et~al.\/}(2009)King, Stellmach, Noir, Hansen \&
  Aurnou]{King2009}
{\sc \au{King, E.M.}, \au{Stellmach, S.}, \au{Noir, J.}, \au{Hansen, U.} \&
  \au{Aurnou, J.~M.}} \yr{2009}  \at{Boundary layer control of rotating
  convection systems}.  \jt{Nature}  \bvol{457},  \pg{301--304}.

\bibitem[Lohse \& Xia(2010)]{Lohse2010}
{\sc \au{Lohse, D.} \& \au{Xia, K.-Q.}} \yr{2010}  \at{Small-scale properties
  of turbulent {Rayleigh-B\'{e}nard} convection}.  \jt{Annu. Rev. Fluid Mech.}
  \bvol{42},  \pg{335--364}.

\bibitem[Ostilla-M\'{o}nico {\em et~al.\/}(2015)Ostilla-M\'{o}nico, Yang,
  van~der Poel, Lohse \& Verzicco]{multigrid2015}
{\sc \au{Ostilla-M\'{o}nico, R.}, \au{Yang, Y.}, \au{van~der Poel, E.~P.},
  \au{Lohse, D.} \& \au{Verzicco, R.}} \yr{2015}  \at{A multiple resolutions
  strategy for direct numerical simulation of scalar turbulence.}  \jt{J.
  Comput. Phys.}  \bvol{301},  \pg{308--321}.

\bibitem[Radko(2013)]{Radko2013}
{\sc \au{Radko, T.}} \yr{2013} {\em Double-diffusive convection\/}.
  \publ{Cambridge, UK: Cambridge University Press}.

\bibitem[Salort {\em et~al.\/}(2014)Salort, Liot, Rusaouen, Seychelles,
  Tisserand, Creyssels, Castaing \& Chillà]{Salort2014}
{\sc \au{Salort, J.}, \au{Liot, O.}, \au{Rusaouen, E.}, \au{Seychelles, F.},
  \au{Tisserand, J.-C.}, \au{Creyssels, M.}, \au{Castaing, B.} \& \au{Chillà,
  F.}} \yr{2014}  \at{Thermal boundary layer near roughnesses in turbulent
  {Rayleigh-B\'{e}nard} convection: Flow structure and multistability}.
  \jt{Phys. Fluids}  \bvol{26},  \pg{015112}.

\bibitem[Shishkina \& Wagner(2011)]{Shishkina2011}
{\sc \au{Shishkina, Olga} \& \au{Wagner, Claus}} \yr{2011}  \at{Modelling the
  influence of wall roughness on heat transfer in thermal convection}.  \jt{J.
  Fluid Mech.}  \bvol{686},  \pg{568?--582}.

\bibitem[Stevens {\em et~al.\/}(2010)Stevens, Verzicco \& Lohse]{Stevens2010}
{\sc \au{Stevens, R.J.A.M.}, \au{Verzicco, R.} \& \au{Lohse, D.}} \yr{2010}
  \at{Radial boundary layer structure and {Nusselt} number in
  {Rayleigh-B\'{e}nard} convection}.  \jt{J. Fluid Mech.}  \bvol{643},
  \pg{495--507}.

\bibitem[Stevens {\em et~al.\/}(2013)Stevens, van~der Poel, Grossmann \&
  Lohse]{GLrefit2013}
{\sc \au{Stevens, R. J. A.~M.}, \au{van~der Poel, E.~P.}, \au{Grossmann, S.} \&
  \au{Lohse, D.}} \yr{2013}  \at{The unifying theory of scaling in thermal
  convection: the updated prefactors}.  \jt{J. Fluid Mech.}  \bvol{730},
  \pg{295--308}.

\bibitem[Turner(1985)]{Turner1985}
{\sc \au{Turner, J.S}} \yr{1985}  \at{Multicomponent convection}.  \jt{Annu.
  Rev. Fluid Mech.}  \bvol{17}~(1),  \pg{11--44}.

\bibitem[Wang {\em et~al.\/}(2017)Wang, Huang \& Xia]{Wang2017}
{\sc \au{Wang, F.}, \au{Huang, S.-D.} \& \au{Xia, K.-Q.}} \yr{2017}
  \at{Thermal convection with mixed thermal boundary conditions: effects of
  insulating lids at the top}.  \jt{J. Fluid Mech.}  \bvol{817},  \pg{R1}.

\bibitem[Wei {\em et~al.\/}(2015)Wei, Weiss \& Ahlers]{Wei2015}
{\sc \au{Wei, P.}, \au{Weiss, S.} \& \au{Ahlers, G.}} \yr{2015}  \at{Multiple
  transitions in rotating turbulent {Rayleigh-B\'{e}nard} convection}.
  \jt{Phys. Rev. Lett.}  \bvol{114},  \pg{114506}.

\bibitem[Xie \& Xia(2017)]{Xie2017}
{\sc \au{Xie, Y.-C.} \& \au{Xia, K.-Q.}} \yr{2017}  \at{Turbulent thermal
  convection over rough plates with varying roughness geometries}.  \jt{J.
  Fluid Mech.}  \bvol{825},  \pg{573--599}.

\bibitem[Yang {\em et~al.\/}(2015)Yang, van~der Poel, Ostilla-M—\'{o}nico,
  Sun, Verzicco, Grossmann \& Lohse]{ddcjfm2015}
{\sc \au{Yang, Y.}, \au{van~der Poel, E.~P.}, \au{Ostilla-M—\'{o}nico, R.},
  \au{Sun, C.}, \au{Verzicco, R.}, \au{Grossmann, S.} \& \au{Lohse, D.}}
  \yr{2015}  \at{Salinity transfer in bounded double diffusive convection}.
  \jt{J. Fluid Mech.}  \bvol{768},  \pg{476--491}.

\bibitem[Yang {\em et~al.\/}(2016)Yang, Verzicco \& Lohse]{ddcjfm2016}
{\sc \au{Yang, Y.}, \au{Verzicco, R.} \& \au{Lohse, D.}} \yr{2016}  \at{Scaling
  laws and flow structures of double diffusive convection in the finger
  regime}.  \jt{J. Fluid Mech.}  \bvol{802},  \pg{667--689}.

\bibitem[Zhu {\em et~al.\/}(2017)Zhu, Stevens, Verzicco \& Lohse]{Zhu2017}
{\sc \au{Zhu, X.}, \au{Stevens, R.J.A.M.}, \au{Verzicco, R.} \& \au{Lohse, D.}}
  \yr{2017}  \at{Roughness-facilitated local 1/2 scaling does not imply the
  onset of the ultimate regime of thermal convection}.  \jt{Phys. Rev. Lett.}
  \bvol{119},  \pg{154501}.

\end{thebibliography}

\end{document}